    \documentclass[twocolumn]{revtex4-2} 
    \usepackage{graphicx}
    \usepackage{subfig}

\graphicspath{{./Figure/}}
\usepackage{dcolumn}
\usepackage{bm}
\usepackage{mathtools}
\usepackage{amsmath,amssymb}
\usepackage{soul}
\usepackage{siunitx}
\usepackage{comment}
\usepackage{url}

\usepackage{calc}
\usepackage{enumerate}
\usepackage{nicefrac}

\newcommand{\bb}[1]{\mathbf{#1}}
\newcommand{\bs}[1]{\boldsymbol{#1}}
\newcommand{\R}{\mathbb R}
\newcommand{\C}{\mathbb C}

\begin{document}

\title[]{Reduced-weight near-cloaks for underwater invisibility}

\author{Davide Enrico Quadrelli}
\affiliation{Department of Mechanical Engineering, Politecnico di Milano, Milan, via La Masa 1, 20156 Milano, Italy}
\author{Sebastiano Cominelli}
\affiliation{Department of Mechanical Engineering, Politecnico di Milano, Milan, via La Masa 1, 20156 Milano, Italy}
\author{Gabriele Cazzulani}
\affiliation{Department of Mechanical Engineering, Politecnico di Milano, Milan, via La Masa 1, 20156 Milano, Italy}
\author{Francesco Braghin}
\affiliation{Department of Mechanical Engineering, Politecnico di Milano, Milan, via La Masa 1, 20156 Milano, Italy}




\email{sebastiano.cominelli@polimi.it}




\begin{abstract}
    Limiting the total weight of an acoustic cloak is of fundamental importance in underwater applications, where buoyancy of the cloaked object is desirable. Unfortunately, it is well known that traditional cloaking strategies imply either  a mass tending to infinity or a total weight equal to the Archimedes' force, thus making a perfect cloak that preserves the buoyancy of the target impossible.
    \\
    In this paper, we discuss strategies to reduce the weight of the cloak seeking a good compromise between weight reduction and acoustic performance. In particular, we compare and combine two existing strategies: the so-called eikonal 
    cloak, where an impedance mismatched cloak is adopted, and the near-cloak, where a non-singular transformation makes the target equivalent to a smaller obstacle.
    We show that properly combining these strategies allows to reduce the mass of the cloak while maintaining a scattering reduction in line with the existing literature. We also investigate radially varying mismatch as a way to further improve the balance between scattering reduction and buoyancy.
\end{abstract}


\maketitle




\section{Introduction}
Acoustic cloaking \cite{chen2010acoustic} refers to the ability to shield an object immersed in a fluid in such a way that it is completely invisible to probing acoustic radiations.
The mathematical tools to achieve this, developed in the early 2000s to reduce the cost of numerical simulations of electromagnetic waves \cite{ward1996refraction}, have been reformulated in the framework of acoustics \cite{cummer2007one, Norris2008, Norris2009} and take the name of Transformation Acoustics.
The key idea is to interpret a given change of coordinates as an effect on the material properties, so that the acoustic space is enriched by a curvature that directs waves as desired. Notably, a singular deformation defines a protected region, guiding waves to propagate around it \cite{Pendry2006, Leonhardt2006}.
Since then, several techniques have been proposed in the literature to achieve acoustic cloaking, such as scattering cancellation \cite{Torrent2008a,sanchez2013experimental}, active cloaking \cite{becker2021broadband}, or other strategies based on inverse design \cite{ahmed2021deterministic,cominelli2022design}.
However, most of these approaches are designed to facilitate the realization and create the desired illusion only for waves of a certain frequency or direction.
Conversely, transformation acoustics allows perfect concealment regardless of the characteristics of the probing wave.
The first strategy, proposed by Cummer and Schurig \cite{cummer2007one}, is known in the literature as Inertial Cloaking (IC) because it relies on anisotropic mass density.
Following this concept, experimental demonstrators have been designed for the so-called carpet cloak \cite{Popa2011,zigoneanu2014three,bi2018experimental}, i.e.\ a device that hides objects close to the ground without using a singular transformation. However, the concealment of an object all-around is still challenging because it requires an infinite mass density, a fact that Norris \cite{Norris2008} called \textit{mass catastrophe}.

Acoustic cloaking is particularly important for underwater applications, where electromagnetic waves are strongly attenuated, but the mass catastrophe is a serious issue since a lightweight device is crucial for buoyancy.
In this scenario, the alternative called elastic or pentamode (PM) \cite{Norris2009} cloaking is pivotal. In this case, the same illusion is achieved by a solid material with isotropic density and whose elastic tensor has 5-over-6 null eigenvalues.
Notably, the density distribution is obtained by rescaling the density of the background fluid by the Jacobian of the transformation, which implies that the total mass of the cloak is equal to that of the undeformed domain.
This means that, regardless of the transformation adopted, the weight of the cloak is equal to the Archimedean buoyancy force generated by the cloak and cloaked object together. This is an improvement with respect to the mass catastrophe, but it is not sufficient since the remaining available thrust for the cloaked object is null, and the system will sink unless the cloaked object has no mass.

The so-called near-cloak strategy \cite{quadrelli2021} is often used to allow a less extreme design of the cloak. In this case, a non-singular transformation is used and the cloak makes the scattering cross section equal to that of a non-zero size target.
A positive side effect is that the weight of the cloak is reduced accordingly, but the performance penalty implied is severe.
As an alternative, impedance-mismatched designs have been proposed: again, a singular transformation makes the target equivalent to a point, but the cloak material properties are scaled by a parameter $\alpha$. In this way, the cloak impedance is scaled accordingly, but the refraction index is maintained and the rays follow the same trajectories. By choosing $\alpha<1$, the weight of the cloak is reduced.
This is known in the literature as eikonal cloak, as a perfect cloaking is recovered in the eikonal limit, provided that the impedance is a continuous function of space \cite{Urzhumov2010,gao2016non}. However, the change of impedance generates undesired reflections, especially for low-frequency waves \cite{cominelli2025non}.
\\
This approach has been applied in various contexts, including acoustics in air \cite{Popa2011} and water \cite{bi2018experimental}, as well as surface water waves \cite{dupont2016cloaking}. However, in many cases, the material properties were altered \textit{ad hoc} to facilitate practical realization rather than to optimize buoyancy together with performance, and no comprehensive analysis is currently available. Furthermore, near-cloak and reduced-weight cloak strategies have been investigated as alternative solutions. This work explores how they can be combined to improve the trade-off between practicality and performance.
We summarize the primary effects of an eikonal cloak and investigate the combined implementation of near-cloak and impedance mismatch. The buoyancy of cloaking devices is discussed, focusing on how to maintain acceptable cloaking performance. Additionally, a general mathematical framework is presented to quantify scattering, based on the acoustic wave solutions for bubbles.
For the sake of brevity, only pure PM cloaks will be considered, since their finite density combined with the possibility of implementing them with non-isotropic solid microstructures makes them more amenable to practical implementation. In any case, the same formulation could easily be extended to the case of ICs.

The manuscript is organized as follows.
After this brief introduction to the problem, §\ref{sec:bubbles} revisits the concept of perfect and near cloaking, illustrating that achieving high performance and minimizing weight are inherently conflicting goals. We then analyze the impact of weight reduction by solving the equivalent problem of bubbles in water.
In §\ref{sec:red wei cloak}, the combined effect of reduced-weight cloaking and near-cloaking is analyzed, discussing the performance in terms of scattered power and available thrust.
In §\ref{sec:alpha variabile}, we let the impedance mismatch vary along the radius to further increase the design degrees of freedom. The
benefit is that the density is scaled less at the water-cloak interface, where the limited mismatch reduces scattering at the interface and allows energy to enter the cloak.    
Finally, we discuss the results and draw the conclusions.

\section{Cloak as a bubble of water in water}
\label{sec:bubbles}

\begin{figure}
    \centering
    \includegraphics[width = 0.5\textwidth]{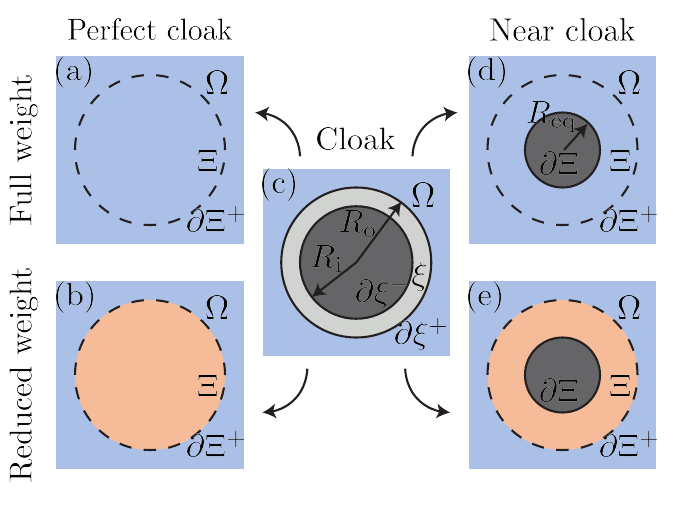}
    \caption{Schematic illustration of the different cloaking strategies for weight reduction.}
    \label{fig:FIG6}
\end{figure}
Figure~\ref{fig:FIG6}a depicts the space $\Xi\cup\Omega=\R^m$ filled with water, where $\Xi$ is a $m$-ball of radius $R_\text{o}$ centered in the origin, and $m\in\{2,3\}$ is the dimension of the problem.
The wave equation
\begin{align}\label{eq:wave eq material}
    \nabla_{\bb X}\cdot\Big(\rho_0^{-1}\nabla_{\bb X}P\Big) &= K_0^{-1} \frac{\partial^2 P}{\partial t^2} 
\end{align}
governs the propagation of a pressure wave $P(\bb X,t)$, where $\rho_0=\qty{1000}{\kilo\gram\per\cubic\meter}$ and ${K_0=\qty{2.2}{\giga\pascal}}$ are the density and bulk modulus of water, respectively. According to acoustics of transformation \cite{cummer2007one}, the map $\bs\chi\colon \bb X\in\Xi \mapsto\bb x\in\xi$ is used to defined the cloak domain $\xi$ shown in Figure~\ref{fig:FIG6}c, and to manipulate \eqref{eq:wave eq material} as
\begin{gather}\label{eq:wave eq spatial}
    \nabla_{\bb x} \cdot \Big(\bs\rho^{-1}\nabla_{\bb x}p\Big) = \frac{1}{K}\frac{\partial^2 p}{\partial t^2},
    \\
    \bs\rho \coloneqq J\rho_0\bb F^{-1}\bb F^{-\top},  \qquad K\coloneqq K_0 J,
\end{gather}
where $\bb F \coloneqq\frac{\partial\bb x}{\partial\bb X}$ is the deformation gradient and $J\coloneqq\det{\bb F}$ is its determinant. The second-order tensor $\bs\rho$ and the scalar field $K$ describe the property distribution of the cloak, and $p(\bb x,t)$ is the pressure field in the deformed domain $\xi$. In this way, the pressure field in $\Omega$ is unchanged and the two scenarios are indistinguishable to a distant observer.
If a singular transformation is chosen, a point of $\Xi$ is mapped onto the inner boundary $\partial\xi^-$ of the cloak, and the cloaked object behaves like a bubble of water embedded in water, producing no scattered wave.
This approach, known as inertial cloak, leads to a perfect concealment but has several criticalities. For instance, Norris \cite{Norris2008} proved that this scenario suffers the so-called \textit{inertia catastrophe}, meaning that an infinite mass is required. In addition, the specific distribution of anisotropic inertia is difficult to obtain in liquids.
\\
In case of the transformation dos not imply rotations (i.e.\ $\bb F$ symmetric) and the left stretch tensor is divergence-free (i.e.\ $\nabla\cdot\bb V=\bs 0$, where $\bb V^2=\bb F^\top\bb F$), \eqref{eq:wave eq spatial} may be rewritten as an elastic wave equation with scalar density $\rho=\rho_0 J^{-1}$, and a five-times singular elasticity tensor $\C$ \cite{Norris2009}, i.e.\ $\C = K_0J^{-1}\,\bb V\otimes\bb V$.
This strategy avoids the \textit{inertia catastrophe}
as the mass of the cloak in $\xi$ is the same as that of the water occupying $\Xi$.
However, it is still unfeasible when targeting buoyancy, because the weight of the bare cloak is exactly equal to the Archimedes' thrust and the cloak-target assembly buoyances only if the target has no weight.

\begin{figure}
    \centering
    \includegraphics[width = .48\textwidth]{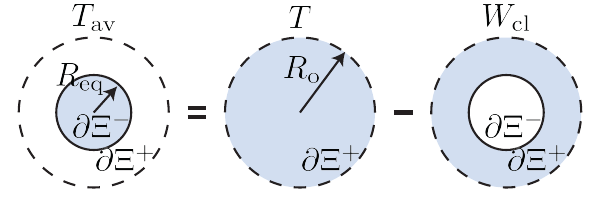}
    \caption{The available thrust $T_\text{av}$ for a near-cloaked target is computed as the difference between the overall thrust $T$ and the weight of the cloak $W_\text{cl}$, and is thus equal to the weight of a volume of water contained in the boundary $\partial\Xi^-$.
    }
    \label{fig:FIG1}
\end{figure}

The situation is slightly improved by adopting the near-cloaking strategy, schematized in Figure~\ref{fig:FIG6}d. With this technique, the inner boundary of the cloak $\partial\xi^-$ is mapped to the boundary $\partial\Xi^-$ of a small obstacle, rather than to a point. Thus, the weight of the cloak $W_{cl}$ is less than the hydrostatic thrust, and the available thrust $T_\text{av}$ is equal to the weight of the water inside $\partial\Xi^-$, as shown in Figure~\ref{fig:FIG1}.
Ideally, the application of a cloak does not change the thrust of the uncloaked object $T_\text{uncloaked}$ so  the ratio $T_\text{av}/T_\text{uncloaked}$ must be close to $1$. For an axisymmetric cloak, it yields
\begin{equation}\label{eq:av thrust}
    \frac{T_\text{av}}{T_\text{uncloaked}}=\left( \frac{R_\text{eq}}{R_\text{i}} \right)^m,\quad m=2,3,
\end{equation}
where $R_\text{eq}$ is the radius of the equivalent object and $R_\text{i}$ the radius of the target to be cloaked.
If one desires to push this ratio as close to one as possible, it is clear that $R_\text{eq}$ should not be much less than $R_\text{i}$, thus implying a limited stealth performance. To gain a qualitative insight, it is sufficient to consider that, in the geometrical acoustics limit, the Total Scattering Cross Section (TSCS) of a rigid circle/sphere that is acoustically illuminated by a plane wave tends to
the actual geometrical cross section, thus the ratio between the TSCS in the cloaked and uncloaked cases is given by:
\begin{equation}
    \frac{TSCS_\text{cloaked}}{TSCS_\text{uncloaked}}
    \to \left( \frac{R_\text{eq}}{R_\text{i}} \right)^{m-1}.
\end{equation}
As a result, the near-cloaking strategy alone cannot simultaneously allow $T_\text{av}/T_\text{uncloaked}$ to be close to one and $TSCS_\text{cloaked}/TSCS_\text{uncloaked}$ to be small.

In the following, we rescale the acoustic impedance of the cloak by a constant value $\alpha<1$ and keep unaltered the sound speed pointwise, so that the weight of the cloak is reduced. In other words,
\begin{equation}\label{eq:c and Z rescaled}
    c=c_0 \qquad Z=\alpha\, Z_0,
\end{equation}
being $c$ the wave speed and $Z$ the acoustic impedance of the cloak, $c_0$ and $Z_0$ refer to water. 
Since the real deformed space $\xi$ and the virtual undeformed space $\Xi$ are equivalent, we will refer to the virtual space for the sake of generality. Any modification of the properties of the virtual space is easily reflected on the real space according to the chosen map $\bs\chi$.

\subsection{Bubble with constant mismatch}

\begin{figure}
    \centering
    \includegraphics[width = .5\textwidth]{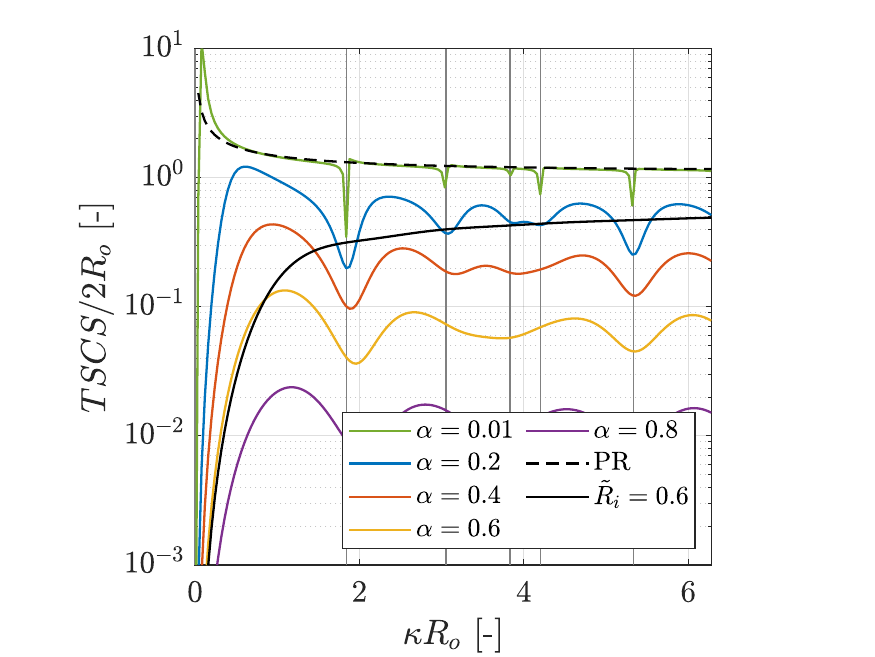}
    \caption{Normalized TSCS for different values of $\alpha$. For increasing contrast the curves tend to that of a pressure release boundary (PR). The solid black line represents the normalized TSCS of a rigid uncloaked target considering $\tilde R_\text{i}=0.6$.
    The vertical lines correspond to the eigenfrequencies of the rigid walled bubble.
    }
    \label{fig:FIG2}
\end{figure}

We call reduced-weight perfect cloak a device obtained by rescaling the perfect-cloak density and elasticity by a factor $\alpha<1$. Such a cloak behaves as a bubble whose impedance contrast is $\alpha$. This configuration is schematized on the left part of Figure~\ref{fig:FIG6}.
\\
In the following, we asses the performance of a cloak considering the TSCS.
We take as reference a circular rigid obstacle with non-dimensional radius $\tilde R_\text{i}\coloneqq R_\text{i}/R_\text{o}=0.6$ because it is close to those adopted in the literature \cite{chen2017broadband,quadrelli2021experimental}. Its TSCS is represented by the black solid line in Figure~\ref{fig:FIG2}. Note that a value of $\tilde R_\text{i}$ close to $1$ increases the cloak performance but complicates its design, as it leads to a zero-thickness cloak.
In contrast, the perfect cloak behaves as a bubble of water in water and has a null TSCS.

We now describe the performance of a reduced-weight perfect cloak considering the bubble in the virtual space $\Xi$ with density $\rho_{\Xi}=\alpha\,\rho_0$ and bulk modulus $K_{\Xi}=\alpha\,K_0$, for an arbitrary $\alpha\in\R^+$. The scattering problem obtained when a plane wave $P_{inc}=\operatorname{Re}(\hat P_{inc}\,e^{i \omega t})=\operatorname{Re}(e^{i \omega t - \bs\kappa\cdot\bb x})$ impinges on this bubble embedded in water is modeled as:
\begin{equation} \label{eq:helmholtz}
\begin{dcases}
    \Delta\hat P_\Omega=-\kappa^2\hat P_\Omega    & \text{in } \Omega\\
    \nabla\cdot\Big(\frac{1}{\alpha}\nabla\hat P_\Xi\Big)=-\frac{\kappa^2}{\alpha}\hat P_\Xi    & \text{in } \Xi\\
    \hat P_\Omega=\hat P_{\Xi}                     &\text{on } \partial\Xi\\
    \frac{\partial\hat P_\Omega}{\partial n}=\frac{1}{\alpha}\frac{\partial\hat P_{\Xi}}{\partial n} & \text{on } \partial\Xi
\end{dcases}
\end{equation}  
where the scattered field $\hat P_\text{sc}=\hat P_\Omega-\hat P_{inc}$ is subjected to the Sommerfeld's radiation condition at infinite. 
The variable $\hat P_{\Xi}$ is used for the field inside the bubble, while $\boldsymbol\kappa$, $\kappa\coloneqq|\bs\kappa|= \omega/c_0$, is the wave vector and $\partial/\partial n$ is the directional derivative normal to $\partial\Xi$. In this and the following section we assume $\alpha$ to be constant, then the second line of~\eqref{eq:helmholtz} is simplified, resembling the first.
In the 2D setting, the solution is found writing the total field as:
\begin{equation}
\label{TotField}
    \hat P_\Omega=\sum_{n=0}^\infty \varepsilon_n i^{3n}J_n(\kappa r)\cos(n\theta)
    +\sum_{n=0}^\infty B_nH_n^{(2)}(\kappa r)\cos(n\theta)
\end{equation}
being $\varepsilon_n=1+\delta_{0n}$, where the incident field is represented according to the Jacobi-Anger expansion and the scattered field is decomposed as a series of vanishing waves. $\delta_{ij}$ is the Kronecker delta. The field inside the bubble is instead expanded in series of Bessel functions of the first kind:
\begin{equation}
    \hat P_{\Xi}=\sum_n C_nJ_n(\kappa r)\cos(n\theta)
\end{equation}
\begin{figure}
    \centering
    \includegraphics[width = .5\textwidth,trim=0 0 350 0,clip]{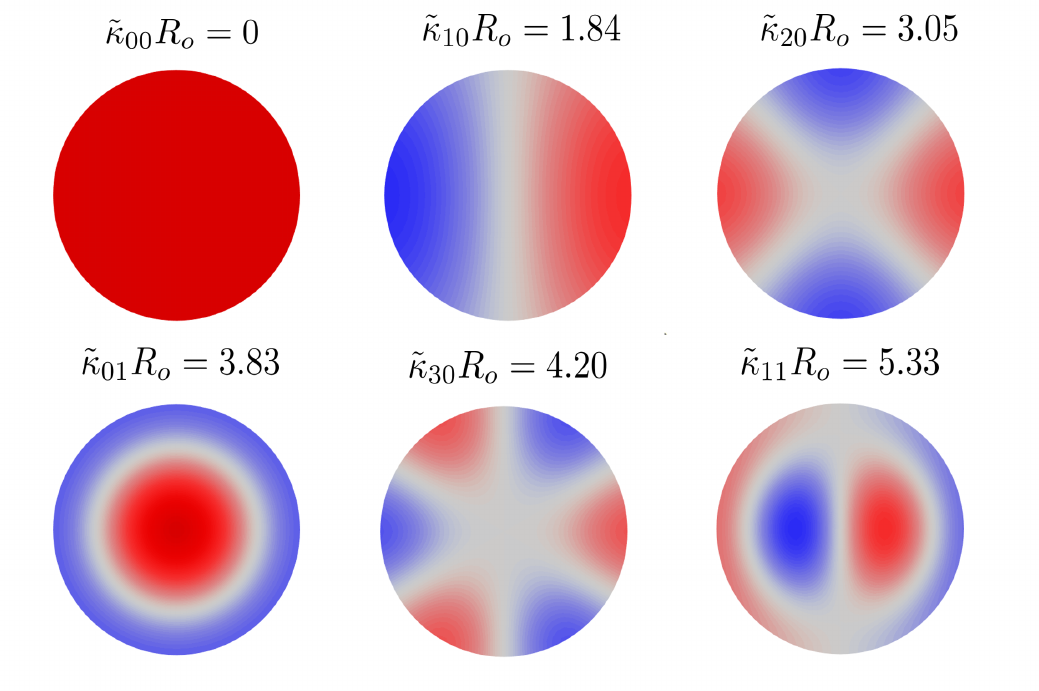}
    \caption{Eigenfrequencies and normal modes for a rigid walled circular bubble.
    }
    \label{fig:FIG4}
\end{figure}
\begin{figure*}
    \centering
    \includegraphics[width = \textwidth]{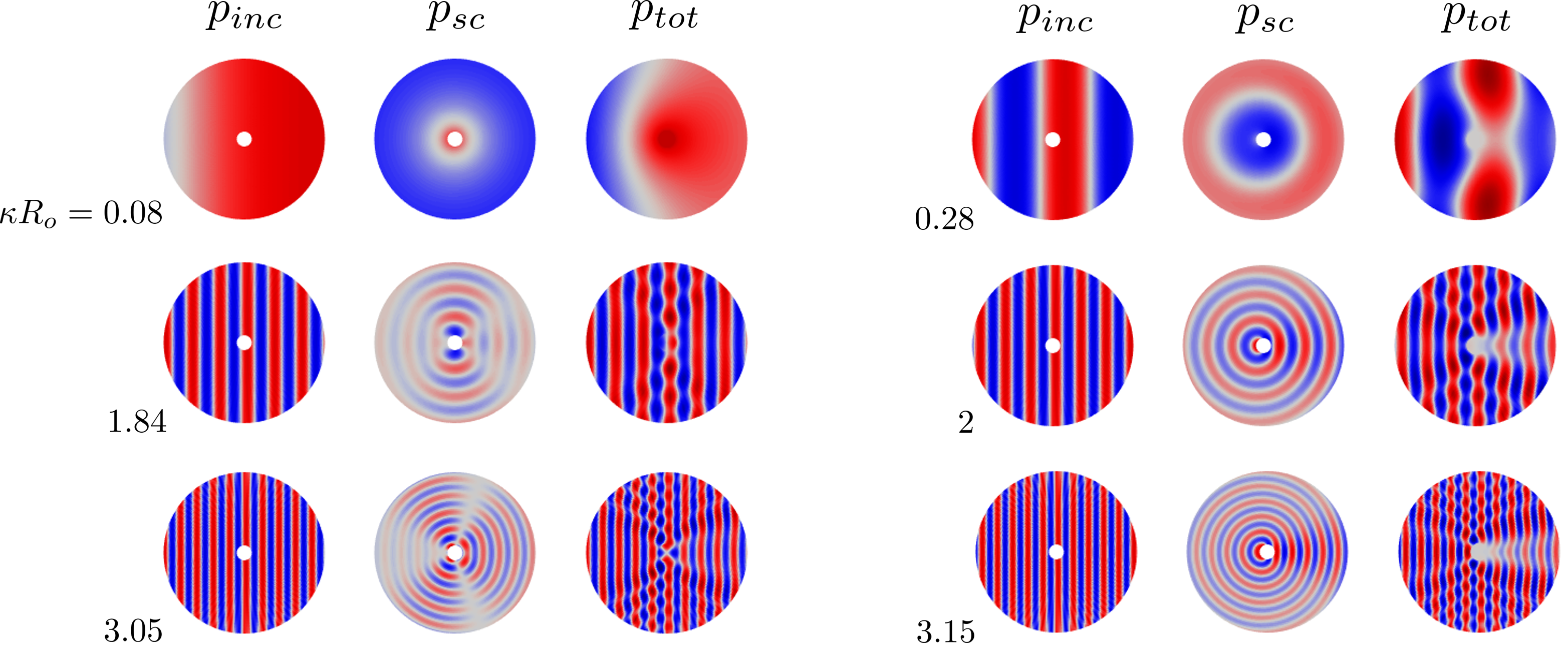}
    \caption{Incident, scattered and total fields calculated for the first three resonances $\kappa R_\text{o}=0.08,\;1.84$ and $3.05$, and for three frequencies close to the resonances but outside the corresponding peak/valleys of TSCS.}
    \label{fig:FIG3}
\end{figure*}
Plugging these definitions into the continuity conditions at the boundary $\partial \Xi$ and exploiting orthogonality, the following system is obtained for the coefficients $B_n$ and $C_n$:
\begin{equation}
\begin{bmatrix}
-H_n^{(2)}(\kappa R_\text{o}) & J_n(\kappa R_\text{o})\\
-\alpha H^{(2) \prime}_n(\kappa R_\text{o}) & J_n^{\prime}(\kappa R_\text{o})
\end{bmatrix}
\begin{bmatrix}
B_n\\
C_n
\end{bmatrix}=
\begin{bmatrix}
\varepsilon_n i^{3n}J_n(\kappa R_\text{o})\\
\alpha \varepsilon_n i^{3n}J_n^{\prime}(\kappa R_\text{o})
\end{bmatrix}
\label{MatriceRisolvente}
\end{equation}
The approach can be easily extended to the three-dimensional case by using spherical Bessel functions and Legendre polynomials. The solution is plotted in Figure~\ref{fig:FIG2} in terms of TSCS normalized with respect to the actual geometric cross section of the bubble $2R_\text{o}$, for varying values of $\alpha$. In the same graph, the dashed black line shows the TSCS of a pressure release boundary having the same geometry of $\partial\Xi^+$, whose scattered field expansion is available from literature \cite{ginsberg2018acoustics}:
\begin{equation}
    B_n=-\frac{\varepsilon_n i^{3n}J_n(\kappa R_\text{o})}{H_n^{(2)}(\kappa R_\text{o})}.
\end{equation}
Figure~\ref{fig:FIG2} compares the performance of the reduced-weight perfect cloak for varying $\alpha$ with the performance of the uncloaked object for ${\tilde R_\text{i}=0.6}$. When $\alpha\approx1$ the TSCS is null, whereas for ${\alpha\rightarrow 0}$ the scatter is dominated by the external size of the cloak $\partial\xi^+$.
The maximum weight reduction $\alpha$ must be evaluated against the comparison with the uncloaked obstacle at hand.
For example, there is no point in using a reduced-weight cloak whose TSCS is higher than that of the uncloaked obstacle while $\alpha\ge0.4$ allows an efficient reduced-weight cloak, for $\tilde R_\text{i}=0.6$.
As a result, $\alpha$ and $\tilde R_\text{i}$ are two design parameters that can be tailored to the designer's need to achieve weight reduction and acoustic performance that meets the requirements at hand.

\begin{figure}
    \centering
    \includegraphics[width = 0.45\textwidth]{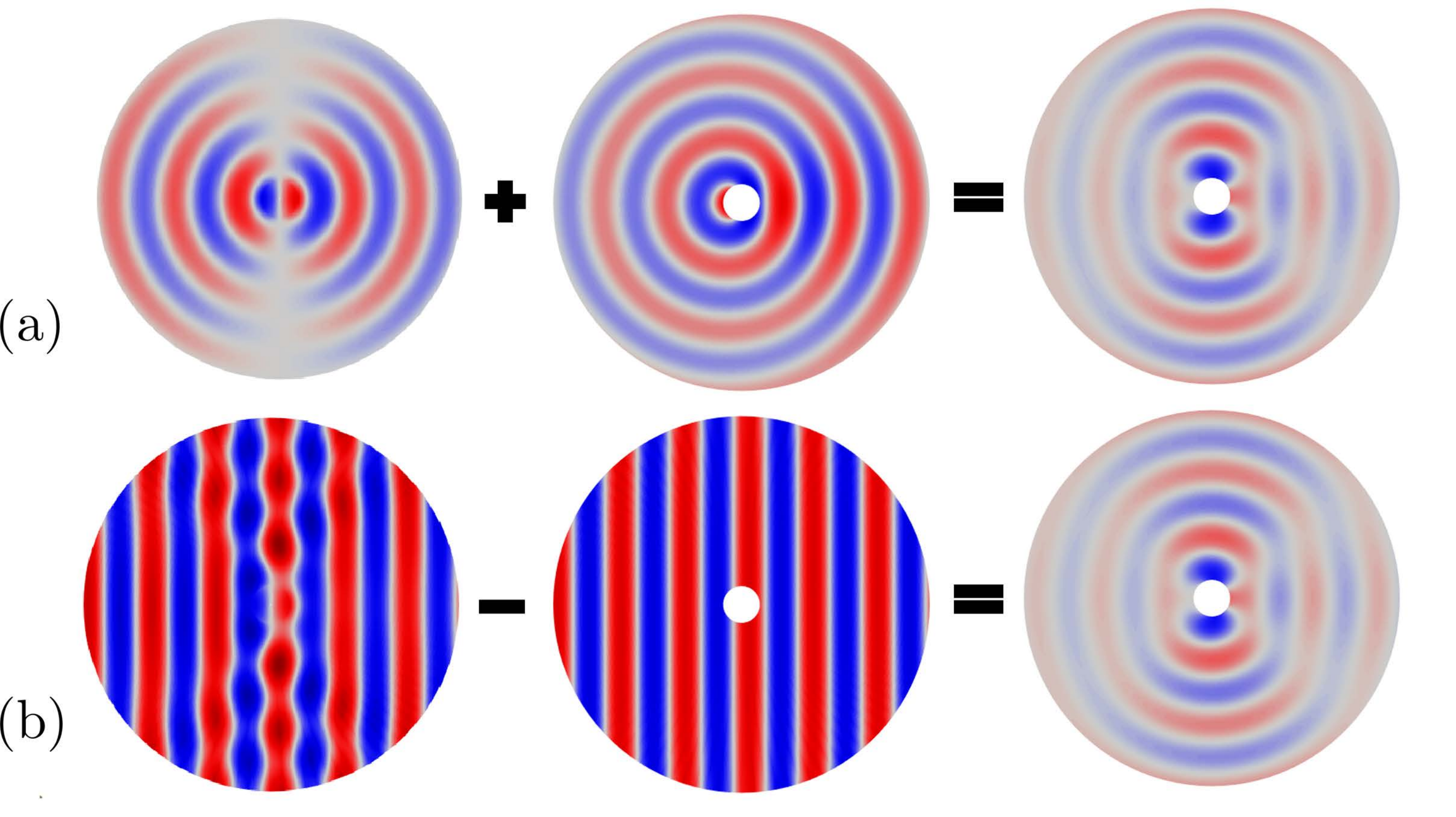}
    \caption{\label{fig:FIG5}{(a)~Superposition between the effect of the resonance and the scattering from a pressure release boundary. (b)~Scattered field obtained as the difference between the total field and the incident one.}}
\end{figure}

The curves approach the line of the pressure release case, as the $\alpha$ parameter tends to zero.
Notably, the TSCS shows only a slight dependence on $\alpha$ at a series of valleys corresponding to the natural frequencies $\tilde\omega_{nm} = c\tilde\kappa_{nm}$ of a rigid-walled bubble \cite{ginsberg2018acoustics}, where $\tilde\kappa_{nm}$ is such that
\begin{equation}
    J_n'(\tilde\kappa_{nm}R_\text{o})=0.
\end{equation}
Figure~\ref{fig:FIG4} shows the eigensolutions of a rigid-walled bubble and their frequency, reported in Figure~\ref{fig:FIG2} as vertical lines.
This correlation is intuitively explained by considering the values of reflection and transmission coefficients for a wave crossing an interface from fluid 1 to fluid 2:
\begin{equation}\label{eq: T and R}
    R=\frac{Z_2-Z_1}{Z_2+Z_1}, \qquad T=\frac{2Z_2}{Z_2+Z_1}.
\end{equation}
In the limit of $\alpha\to0$, when a wave propagates from the outside to the inside of the bubble, the reflection coefficient tends to unity and the transmission to zero. This implies that the energy cannot effectively penetrate the boundary and the acoustic fields inside and outside $\Xi$ are decoupled. In the opposite direction, i.e.\ the soft-to-hard propagation, the reflection coefficient tends to unity, while $T\to2$. This means that the radiation penetrating the bubble, even if it is small, excites the modes of the cavity and is then easily transmitted to the outside, affecting the total scattered field.
\\
This intuition is supported by inspection of the system response shown in Figure~\ref{fig:FIG3} where incident, scattered, and total fields are reported for $\alpha=0.01$ and $\kappa R_\text{o}=0.08$, $1.84$, and $3.05$, i.e.\ the first three resonances. Looking at the total and scattered fields, we appreciate that monopolar, dipolar and quadrupolar distributions arise inside the bubble, resembling the mode shape depicted in Figure~\ref{fig:FIG4}.
We compare these responses with those at $\kappa R_\text{o}= 0.28$, $2$ and $3.15$, chosen close to the previous three but outside the peak/valleys. In these cases, the pressure inside $\Xi$ is close to zero.
\\
We prove that this phenomenon is advantageous by looking back to~\eqref{MatriceRisolvente}. For $\alpha\rightarrow0$, the second line of the system becomes identical to the characteristic equation for the eigenproblem of the rigid-walled bubble. Thus, $C_n\rightarrow0$ when $\kappa\neq\tilde{\kappa}_{nm}$, and the first equation reduces to the pressure release solution.
For $\kappa=\tilde\kappa_{nm}$, the first equation gives:
\begin{equation}
    B_n\approx\frac{C_nJ_n(\tilde{\kappa}_{nm} R_\text{o})}{H_n^{(2)}(\tilde{\kappa}_{nm} R_\text{o})}-\frac{\varepsilon_ni^{3n}J_n(\tilde{\kappa}_{nm} R_\text{o})}{H_n^{(2)}(\tilde{\kappa}_{nm} R_\text{o})}
\end{equation}
which allows for interpretation of the solution as a superposition of the scattered field from a pressure release boundary and the field obtained imposing on $\partial\Xi$ the pressure distribution of the corresponding natural mode of the rigid-walled bubble.
\\
In Figure~\ref{fig:FIG5}a such a superposition is computed for the valley at $\kappa R_\text{o}=1.84$ (dipolar resonance). The result is compared to the scattered field in Figure~\ref{fig:FIG5}b obtained subtracting the incident field to the total field computed via~\eqref{MatriceRisolvente}.
The previous analysis suggests that, apart from the breathing mode at $\kappa R_\text{o}\approx0$, known as the Minnaert resonance \cite{ammari2018minnaert}, the interference between the scattering due to the impedance mismatch and the radiated field from the resonance is destructive, limits the overall TSCS of the bubble. Therefore, the performance of a reduced-weight cloak benefits from this.

\section{The reduced-weight near cloak}
\label{sec:red wei cloak}

We now assess the acoustic performance of what we call \textit{reduced-weight near-cloaking}. The scattering problem in the material frame $\bb X$ is
\begin{equation}\label{eq:helm con ostacolo}
\begin{dcases}
    \Delta\hat P_\Omega=-\kappa^2\hat P_\Omega    & \text{in } \Omega \\
    \nabla\cdot\Big(\frac{1}{\alpha}\nabla\hat P_\Xi\Big)=-\frac{\kappa^2}{\alpha}\hat P_\Xi    & \text{in } \Xi \\
    \hat P_\Omega=\hat P_{\Xi}                     & \text{on } \partial\Xi^+ \\
    \frac{\partial\hat P_\Omega}{\partial n}=\frac{1}{\alpha}\frac{\partial\hat P_\Xi}{\partial n} & \text{on } \partial\Xi^+ \\
    \displaystyle\frac{\partial\hat P_{\Xi}}{\partial n}=0 & \text{on } \partial\Gamma
\end{dcases}
\end{equation}
where a rigid obstacle is assumed inside the boundary $\Gamma$.
The field $\hat P_\Omega$ is expanded as in~\eqref{TotField}, while the field $\hat P_{\Xi}$ in the annular region $\Xi$ is written as
\begin{equation}
    \hat P_{\Xi}=\sum_n \left[ C_nJ_n(\kappa r) +D_nH_n^{(2)}(\kappa r) \right]\cos(n\theta)
\end{equation}
Upon substitution, the following system of equations is found:
\begin{multline}
\label{eq:15}
    \begin{bmatrix}
    0                 & J_n^\prime(\kappa R_\text{eq})  &  H_n^{(2)\prime}(\kappa R_\text{eq})\\
    -H_n^{(2)}(\kappa R_\text{o})      & J_n(\kappa R_\text{o})   & H_n^{(2)}(\kappa R_\text{o})\\
    -H_n^{(2)\prime}(\kappa R_\text{o}) & J_n^\prime(\kappa R_\text{o})/\alpha &  H_n^{(2)\prime}(\kappa R_\text{o})/\alpha\\
    \end{bmatrix}
    \begin{bmatrix}
    B_n  \\    C_n \\    D_n
    \end{bmatrix} 
    \\ =
    \begin{bmatrix}
    0  \\   \varepsilon_n i^{3n}J_n(\kappa R_\text{o})  \\    \varepsilon_n i^{3n}J_n^\prime(\kappa R_\text{o})
    \end{bmatrix},
\end{multline}
being $R_\text{o}$ and $R_\text{eq}$ the radii of $\partial\Xi^+$ and $\partial\Xi^-$, respectively.

\subsection{Near cloak}

\begin{figure}
    \centering
    \includegraphics[width=0.5\textwidth]{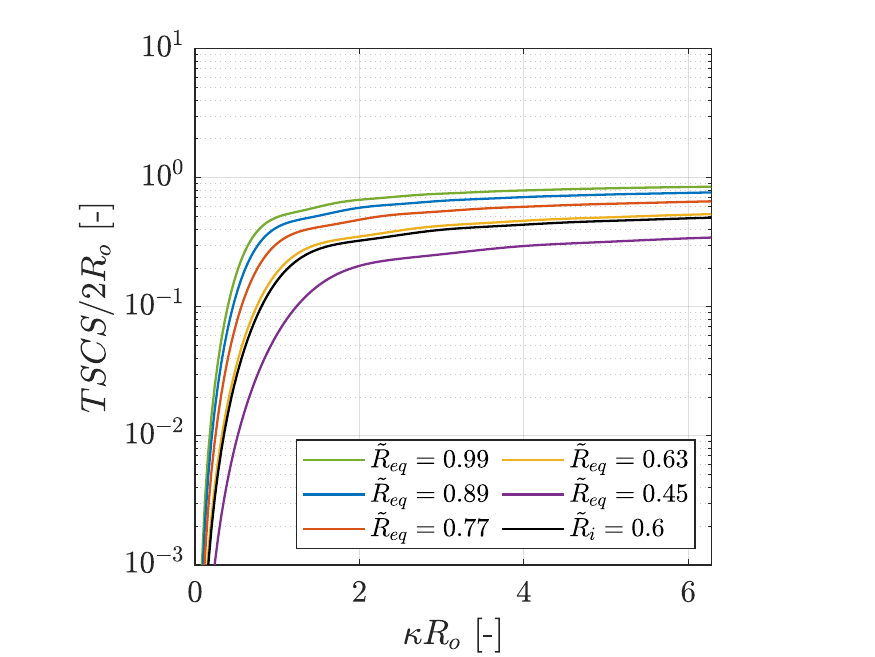}
    \caption{Normalised TSCS of a near-cloaked obstacle for varying $\tilde R_\text{eq}$.
    }
    \label{fig:near_cloak}
\end{figure}

When $\alpha=1$, the configuration simplifies to the near cloak described in §\ref{sec:bubbles}.
In this case, the mass of the full-weight cloak is $m_\text{FW}=\pi(R_\text{o}^2-R_\text{eq}^2)$, depending on the radius of the equivalent obstacle. Using the reduced-weight strategy instead, the mass of the cloak is $m_\text{RW}=\alpha\,\pi R_\text{o}^2$.
Thus, the same mass is obtained in the two scenarios when $R_\text{eq} = R_\text{o}\sqrt{1-\alpha}$. Table~\ref{tab:alpha vs R_eq} compares the values of $\alpha$ considered before with the radii $\tilde R_\text{eq}\coloneqq R_\text{eq}/R_\text{o}$ that provide the same mass.
\\
The TSCS of the near cloak is shown in Figure~\ref{fig:near_cloak} for these values of $\tilde R_\text{eq}$.
Clearly, only when $\tilde R_\text{eq}<\tilde R_\text{i}$ a TSCS smaller than that of the bare obstacle is achieved.
The value typically used in the literature \cite{chen2017broadband,quadrelli2021experimental} is $\tilde R_\text{eq}\approx0.25$.

\begin{table}[]
    \centering
    $\begin{array}{c|ccccc}
    \alpha \,\,\,\,\,[-] & 0.01 & 0.2 & 0.4 & 0.6 & 0.8
    \\
    \tilde R_\text{eq} [-]& 0.99 & 0.894 & 0.775 & 0.633  & 0.447
    \end{array}$
    \caption{The same mass of the cloak is obtained either using a reduced-weight cloak with impedance mismatch $\alpha$ or a near cloak with an equivalent radius $\tilde R_\text{eq}$.}
    \label{tab:alpha vs R_eq}
\end{table}

\subsection{Reduced-weight near cloak}

\begin{figure}
\centering
    \includegraphics[trim={0 0 0 0},width=0.5\textwidth]{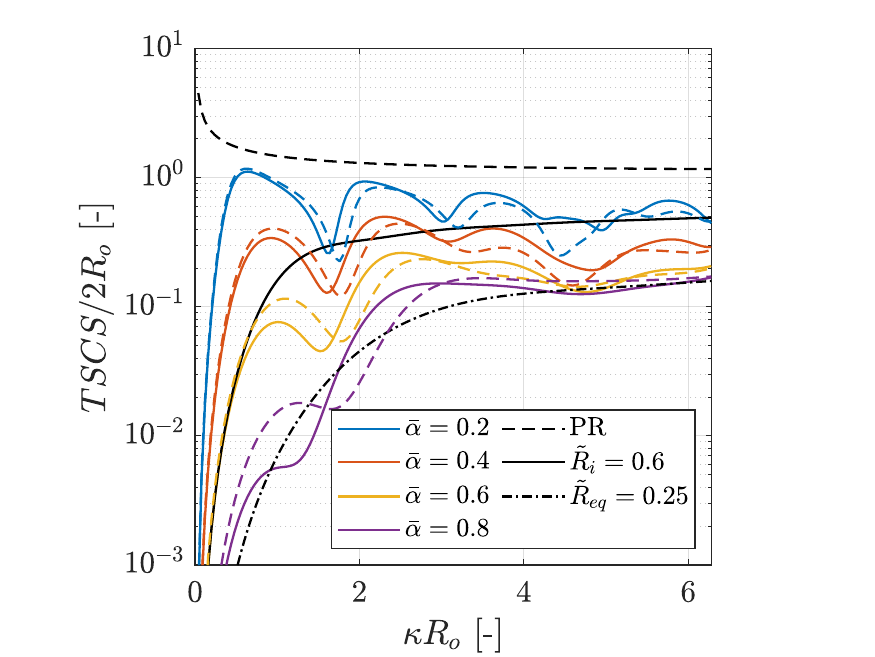}
    \caption{TSCS compared between the constant mismatch (solid line) and varying mismatch (dashed line). The dash-dotted line represents the TSCS of the virtual obstacle when $\tilde R_\text{eq}=0.25$.
    }
    \label{fig:TSCS rif}
\end{figure}

In search of better performance, we combine the two strategies.
We choose $\tilde R_\text{eq}=0.25$ and compute the TSCS of the reduced-weight near cloak, as shown in Figure~\ref{fig:TSCS rif}.
A value of $\alpha$ equal to $=0.8$ or higher allows a performance in line with those achieved in literature and with a total mass reduced to
\begin{equation}
    \alpha (1-\tilde R_\text{eq}^2) = 0.8 \cdot 0.938 = \qty{75}{\percent}
\end{equation}
of the full-weight perfect cloak.
Note that the larger contribution comes from the impedance mismatch.

\begin{figure}
    \centering
    \includegraphics[width = 0.5\textwidth]{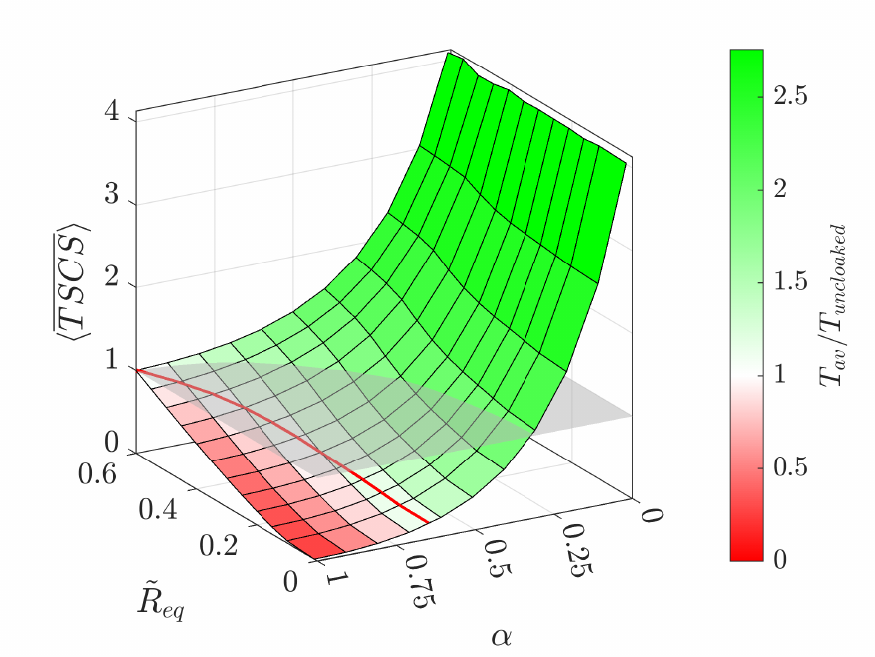}
    \caption{Performance of a reduced-weight near-cloak: the horizontal axes indicate $\alpha$ and $\tilde R_\text{eq}$, the vertical axis $\langle \overline{\text{TSCS}}\rangle$, and the color scale indicates the hydrostatic thrust.
    }
    \label{fig:FIG9}
\end{figure}

The performance of combining the two strategies is summarized in Figure~\ref{fig:FIG9}: the value $\langle{\text{TSCS}}\rangle$ is computed as the frequency average of TSCS in the interval $\kappa R_\text{o}\in[0,2\pi]$. The value displayed in the figure is $\langle\overline{\text{TSCS}}\rangle\coloneqq \langle\text{TSCS}_\text{cloaked}\rangle/ \langle\text{TSCS}_\text{uncloaked}\rangle$.
A horizontal plane at unitary height shows the region where the mean TSCS is reduced by the cloak. The color represents instead the ratio between the available thrust $T_\text{av}=\rho_0g\,\alpha\,\pi(R_\text{o}^2-R_\text{eq}^2)$ and the hydrostatic force in the uncloaked scenario $T_\text{uncloaked}=\rho_0g\,\alpha\,\pi R_\text{i}^2$, where $g$ is the acceleration of gravity.
The red line is the isoline $T_\text{av}/T_\text{uncloaked} =1$, where the lift force felt by the target is unchanged from the uncloaked to the cloaked scenarios.
This line divides the surface into a red and a green region. The former represents situations where the cloak adds more weight than Archimedes' force. The latter represents designs where the additional Archimedes' force is greater than the weight of the cloak, resulting in a buoyancy force greater than the uncloaked case.
\\
The full-weight near-cloak scenario is obtained with $\alpha=1$ and for decreasing values of $\tilde R_\text{eq}$. In this case the scattered field is nearly proportional to the equivalent radius $R_\text{eq}$. At the same time, however, the loss of available thrust is important even for low reductions of TSCS, thus limiting the minimum practical $\tilde R_\text{eq}$.
Conversely, the reduced-weight near-cloak strategy allows to go lower in $\tilde R_\text{eq}$ and to reduce the overall weight by considering $\alpha<1$, thus reaching a balance between the added weight and the increase in buoyancy force.

\section{Radially varying mismatch}
\label{sec:alpha variabile}
In the previous sections, we have shown that a mismatched cloak guides waves similarly to the ideal case, but with additional reflections at the interface with the background. Since the cloak is inherently inhomogeneous, with properties varying along its radius, a uniform rescaling of the density is unlikely to be optimal.
In this section, we investigate the effects of spatially varying mismatch. This approach introduces an additional degree of freedom in the design and allows for a better performance given the desired mass reduction by smoothly varying the impedance along the cloak radius. In general, a varying mismatch may introduce \cite{cominelli2025non}:
\begin{description}
    \item[Interface reflections] discontinuities at the interfaces $\partial\Xi^+$ and $\partial\Xi^-$ cause reflections, similar to the constant mismatch case analyzed earlier.
    \item[Bulk dispersion] variations in impedance within the domain $\Xi$ may alter the sound speed with frequency, effectively introducing dispersion.
\end{description}
Both effects significantly impact the performance of the cloak and require careful consideration during the design process.

\subsection{Bubble with graded mismatch}
\begin{figure}
\centering  
    \includegraphics[trim=0 0 0 0,width=.5\textwidth]{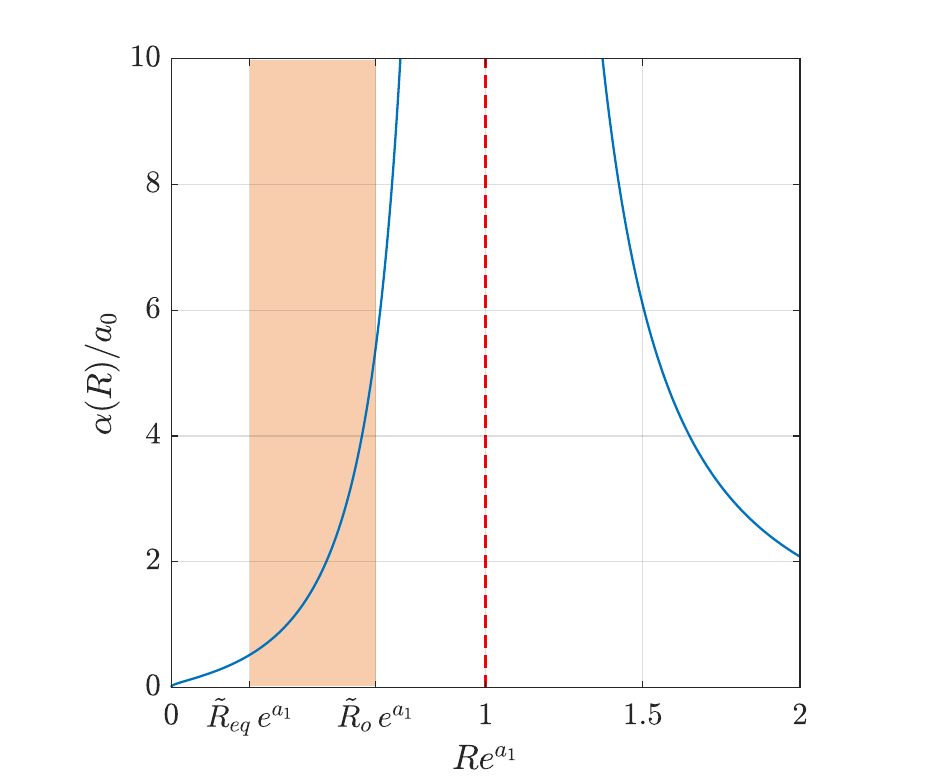}
    \caption{impedance profile that prevents dispersion in the cloak; all points in the domain $\Xi$ have a radius within the interval highlighted in orange. 
    }
    \label{fig:fam f}
\end{figure}

Typically, cloaks are designed to operate at wavelengths similar to or smaller than the size of the cloaked object. A smooth impedance transition has been shown to be particularly beneficial for waves in such a frequency range \cite{cominelli2025non}. In particular, an impedance profile has been derived to reduce dispersion.
\\
As before, we consider the virtual reference frame and modify the cloak properties using the undeformed coordinates. The actual properties of the cloak may be obtained through the map $\bs\chi$. According to Cominelli \cite{cominelli2025non}, the only function that changes the impedance of an axisymmetric lens without introducing dispersion is $\alpha(R)=a_0(a_1+N(R))^{-2}$, where $N(R)=\int_{R_\text{eq}}^{R} n(\tilde R)/\tilde R\,\text d\tilde R$, and $a_0,a_1\in\R$ are arbitrary constants. Here, $n$ represents the refractive index, which in this case is $1$, since the virtual domain $\Xi$ simply contains water. Thus, we change the impedance along the cloak radius as
\begin{equation}
    \alpha(R) = a_0\Big(a_1+\ln\frac{R}{R_\text{eq}}\Big)^{-2}.
\end{equation}
Figure~\ref{fig:fam f} shows the plot of $\alpha(R)$ normalized with respect to $a_0$ and $a_1$. Note that the function can be either increasing or decreasing, depending on the choice of the coefficient $a_1$. To reduce the mass and limit reflections, we choose it so that the impedance profile has a maximum in correspondence of the outer radius, in contact with water, and a minimum at the inner radius.
The opposite can be beneficial when a cloak is designed for a soft and light medium, such as air \cite{Popa2011,zigoneanu2014three}.

Given a desired average mass reduction $\bar\alpha\in\R$, we set the change in impedance between the outer and inner surfaces of the cloak $\bar\mu\in\R$.
Then, the two coefficients $a_0$, $a_1$ are calculated by solving the following equations
\begin{equation}\label{eq:2 conds}
    \begin{dcases}
    m_\text{RW} = \int_0^{2\pi}\int_{R_\text{eq}}^{R_\text{o}} \rho_0\,\alpha(R) R\,\text dR\,\text d\theta = \bar\alpha\, m_\text{FW}
    \\
    \alpha(R_\text{eq}) = \bar\mu\, \alpha(R_\text{o})
    \end{dcases},
\end{equation}
where $m_\text{RW}$ and $m_\text{FW}=\rho_0\,\pi(R_\text{o}^2-R_\text{eq}^2)$ are the masses of the reduced-weight and full-weight cloaks, respectively.
If $\bar\mu=1$ there is no reflection in the eikonal limit, since only the jump of the impedance over $\partial\Xi^+$ is relevant, not any change of the slope \cite{cominelli2025non}. However, this choice together with \eqref{eq:2 conds}$_1$ would imply extreme properties at $R=\tilde R_\text{eq}$, i.e.\ very small density.
As an example, we set $\bar\mu=0.5$ and $\bar\alpha\in\{0.2,0.4,0.6,0.8\}$ and consider $\tilde R_\text{eq}=0.25$.
From \eqref{eq:2 conds}$_2$ we have $a_1=3.347$ thus, from \eqref{eq:2 conds}$_1$ we obtain $a_0=13.788 \,\bar\alpha$.
The TSCS is computed in these four scenarios and displayed in Figure~\ref{fig:TSCS rif} where it is compared with a non-cloaked target with $\tilde R_\text{i}=0.6$.
The advantage of a varying impedance is that at high frequencies ($\kappa R_\text{o} > 2$, i.e., $\lambda < \pi R_\text{o}$, a common scenario in practical applications) it effectively reduces reflections. 
In the eikonal limit, only the impedance jump at $\partial\Xi^+$ affects the performance \cite{cominelli2025non} and, since for a given mass reduction such a jump is smaller than that obtained with a constant mismatch, the cloak has a higher performance.

\section{Conclusion}
Perfect acoustic cloaking of underwater targets causes the cloak-target assembly to sink unless the target has no mass. In this paper, we analyze the trade-off between buoyancy and acoustic performance of pentamode cloaks and explore the implications of this trade-off for practical implementations.
\\
We show that the design space for buoyant cloaks can be expanded by employing a near-cloaking technique with simultaneous weight reduction, achieved by rescaling both density and acoustic impedance by a common factor $\alpha<1$. In particular, the total scattering cross section increases due to reflections at the mismatched surface of the cloak, but it is mitigated by the resonances occurring inside the cloak. This makes the increase in scattered power due to a decrease in $\alpha$ slower in certain frequency ranges than in others. The extent and width of these frequency ranges can be tailored by varying $\alpha$.
\\
In addition, we have incorporated a graded impedance profile that smooths transitions and minimizes scattering caused by abrupt impedance changes. This approach maintains the buoyancy of the cloak and improves its performance, especially at high frequency.
\\
The increased degrees of freedom offered by the combination of $\alpha$ and $\tilde R_\text{eq}$ provide designers with a versatile framework for optimizing the trade-off between buoyancy and invisibility, thus advancing the practical implementation of cloaking concepts in underwater acoustics.

\section*{Conflict of interest}
All authors declare that they have no conflicts of interest.

\section*{Data availability}
The codes used in this paper are available upon request.

\begin{acknowledgments}
    The authors are grateful to the Italian Ministry of Defense for the financial support granted to this work through the PNRM ``SUWIMM''.
\end{acknowledgments}





\bibliography{Mybibliography}

\begin{thebibliography}{24}%
\makeatletter
\providecommand \@ifxundefined [1]{%
 \@ifx{#1\undefined}
}%
\providecommand \@ifnum [1]{%
 \ifnum #1\expandafter \@firstoftwo
 \else \expandafter \@secondoftwo
 \fi
}%
\providecommand \@ifx [1]{%
 \ifx #1\expandafter \@firstoftwo
 \else \expandafter \@secondoftwo
 \fi
}%
\providecommand \natexlab [1]{#1}%
\providecommand \enquote  [1]{``#1''}%
\providecommand \bibnamefont  [1]{#1}%
\providecommand \bibfnamefont [1]{#1}%
\providecommand \citenamefont [1]{#1}%
\providecommand \href@noop [0]{\@secondoftwo}%
\providecommand \href [0]{\begingroup \@sanitize@url \@href}%
\providecommand \@href[1]{\@@startlink{#1}\@@href}%
\providecommand \@@href[1]{\endgroup#1\@@endlink}%
\providecommand \@sanitize@url [0]{\catcode `\\12\catcode `\$12\catcode `\&12\catcode `\#12\catcode `\^12\catcode `\_12\catcode `\%12\relax}%
\providecommand \@@startlink[1]{}%
\providecommand \@@endlink[0]{}%
\providecommand \url  [0]{\begingroup\@sanitize@url \@url }%
\providecommand \@url [1]{\endgroup\@href {#1}{\urlprefix }}%
\providecommand \urlprefix  [0]{URL }%
\providecommand \Eprint [0]{\href }%
\providecommand \doibase [0]{https://doi.org/}%
\providecommand \selectlanguage [0]{\@gobble}%
\providecommand \bibinfo  [0]{\@secondoftwo}%
\providecommand \bibfield  [0]{\@secondoftwo}%
\providecommand \translation [1]{[#1]}%
\providecommand \BibitemOpen [0]{}%
\providecommand \bibitemStop [0]{}%
\providecommand \bibitemNoStop [0]{.\EOS\space}%
\providecommand \EOS [0]{\spacefactor3000\relax}%
\providecommand \BibitemShut  [1]{\csname bibitem#1\endcsname}%
\let\auto@bib@innerbib\@empty
\bibitem [{\citenamefont {Chen}\ and\ \citenamefont {Chan}(2010)}]{chen2010acoustic}%
  \BibitemOpen
  \bibfield  {author} {\bibinfo {author} {\bibfnamefont {H.}~\bibnamefont {Chen}}\ and\ \bibinfo {author} {\bibfnamefont {C.~T.}\ \bibnamefont {Chan}},\ }\bibfield  {title} {\bibinfo {title} {Acoustic cloaking and transformation acoustics},\ }\href@noop {} {\bibfield  {journal} {\bibinfo  {journal} {Journal of Physics D: Applied Physics}\ }\textbf {\bibinfo {volume} {43}},\ \bibinfo {pages} {113001} (\bibinfo {year} {2010})}\BibitemShut {NoStop}%
\bibitem [{\citenamefont {Ward}\ and\ \citenamefont {Pendry}(1996)}]{ward1996refraction}%
  \BibitemOpen
  \bibfield  {author} {\bibinfo {author} {\bibfnamefont {A.}~\bibnamefont {Ward}}\ and\ \bibinfo {author} {\bibfnamefont {J.~B.}\ \bibnamefont {Pendry}},\ }\bibfield  {title} {\bibinfo {title} {Refraction and geometry in maxwell's equations},\ }\href@noop {} {\bibfield  {journal} {\bibinfo  {journal} {Journal of modern optics}\ }\textbf {\bibinfo {volume} {43}},\ \bibinfo {pages} {773} (\bibinfo {year} {1996})}\BibitemShut {NoStop}%
\bibitem [{\citenamefont {Cummer}\ and\ \citenamefont {Schurig}(2007)}]{cummer2007one}%
  \BibitemOpen
  \bibfield  {author} {\bibinfo {author} {\bibfnamefont {S.~A.}\ \bibnamefont {Cummer}}\ and\ \bibinfo {author} {\bibfnamefont {D.}~\bibnamefont {Schurig}},\ }\bibfield  {title} {\bibinfo {title} {One path to acoustic cloaking},\ }\href@noop {} {\bibfield  {journal} {\bibinfo  {journal} {New Journal of Physics}\ }\textbf {\bibinfo {volume} {9}},\ \bibinfo {pages} {45} (\bibinfo {year} {2007})}\BibitemShut {NoStop}%
\bibitem [{\citenamefont {Norris}(2008)}]{Norris2008}%
  \BibitemOpen
  \bibfield  {author} {\bibinfo {author} {\bibfnamefont {A.~N.}\ \bibnamefont {Norris}},\ }\bibfield  {title} {\bibinfo {title} {Acoustic cloaking theory},\ }\href@noop {} {\bibfield  {journal} {\bibinfo  {journal} {Proceedings of the Royal Society A}\ }\textbf {\bibinfo {volume} {464}},\ \bibinfo {pages} {2411} (\bibinfo {year} {2008})}\BibitemShut {NoStop}%
\bibitem [{\citenamefont {Norris}(2009)}]{Norris2009}%
  \BibitemOpen
  \bibfield  {author} {\bibinfo {author} {\bibfnamefont {A.~N.}\ \bibnamefont {Norris}},\ }\bibfield  {title} {\bibinfo {title} {Acoustic metafluids},\ }\href@noop {} {\bibfield  {journal} {\bibinfo  {journal} {J. Acoust. Soc. Am.}\ }\textbf {\bibinfo {volume} {125}},\ \bibinfo {pages} {839} (\bibinfo {year} {2009})}\BibitemShut {NoStop}%
\bibitem [{\citenamefont {Pendry}\ \emph {et~al.}(2006)\citenamefont {Pendry}, \citenamefont {Schurig},\ and\ \citenamefont {Smith}}]{Pendry2006}%
  \BibitemOpen
  \bibfield  {author} {\bibinfo {author} {\bibfnamefont {J.~B.}\ \bibnamefont {Pendry}}, \bibinfo {author} {\bibfnamefont {D.}~\bibnamefont {Schurig}},\ and\ \bibinfo {author} {\bibfnamefont {D.~R.}\ \bibnamefont {Smith}},\ }\bibfield  {title} {\bibinfo {title} {Controlling electromagnetic fields},\ }\href@noop {} {\bibfield  {journal} {\bibinfo  {journal} {Science}\ }\textbf {\bibinfo {volume} {312}},\ \bibinfo {pages} {1780} (\bibinfo {year} {2006})}\BibitemShut {NoStop}%
\bibitem [{\citenamefont {Leonhardt}(2006)}]{Leonhardt2006}%
  \BibitemOpen
  \bibfield  {author} {\bibinfo {author} {\bibfnamefont {U.}~\bibnamefont {Leonhardt}},\ }\bibfield  {title} {\bibinfo {title} {Optical conformal mapping},\ }\href@noop {} {\bibfield  {journal} {\bibinfo  {journal} {Science}\ }\textbf {\bibinfo {volume} {312}},\ \bibinfo {pages} {1777} (\bibinfo {year} {2006})}\BibitemShut {NoStop}%
\bibitem [{\citenamefont {Torrent}\ and\ \citenamefont {Sánchez-Dehesa}(2008)}]{Torrent2008a}%
  \BibitemOpen
  \bibfield  {author} {\bibinfo {author} {\bibfnamefont {D.}~\bibnamefont {Torrent}}\ and\ \bibinfo {author} {\bibfnamefont {J.}~\bibnamefont {Sánchez-Dehesa}},\ }\bibfield  {title} {\bibinfo {title} {Acoustic cloaking in two dimensions: a feasible approach},\ }\href@noop {} {\bibfield  {journal} {\bibinfo  {journal} {New J. Phys.}\ }\textbf {\bibinfo {volume} {10}},\ \bibinfo {pages} {063015} (\bibinfo {year} {2008})}\BibitemShut {NoStop}%
\bibitem [{\citenamefont {Sanchez-Dehesa}\ \emph {et~al.}(2013)\citenamefont {Sanchez-Dehesa}, \citenamefont {Garc{\'\i}a-Chocano}, \citenamefont {Climente}, \citenamefont {Cervera}, \citenamefont {Gomez-Lozano}, \citenamefont {Sanchis}, \citenamefont {LLopis-Pontiveros},\ and\ \citenamefont {Juan}}]{sanchez2013experimental}%
  \BibitemOpen
  \bibfield  {author} {\bibinfo {author} {\bibfnamefont {J.}~\bibnamefont {Sanchez-Dehesa}}, \bibinfo {author} {\bibfnamefont {V.~M.}\ \bibnamefont {Garc{\'\i}a-Chocano}}, \bibinfo {author} {\bibfnamefont {A.}~\bibnamefont {Climente}}, \bibinfo {author} {\bibfnamefont {F.}~\bibnamefont {Cervera}}, \bibinfo {author} {\bibfnamefont {V.}~\bibnamefont {Gomez-Lozano}}, \bibinfo {author} {\bibfnamefont {L.}~\bibnamefont {Sanchis}}, \bibinfo {author} {\bibfnamefont {R.}~\bibnamefont {LLopis-Pontiveros}},\ and\ \bibinfo {author} {\bibfnamefont {M.-P.}\ \bibnamefont {Juan}},\ }\bibfield  {title} {\bibinfo {title} {Experimental demonstration of a three-dimensional acoustic cloak based on a cancelation effect},\ }\href@noop {} {\bibfield  {journal} {\bibinfo  {journal} {The Journal of the Acoustical Society of America}\ }\textbf {\bibinfo {volume} {133}},\ \bibinfo {pages} {3374} (\bibinfo {year} {2013})}\BibitemShut {NoStop}%
\bibitem [{\citenamefont {Becker}\ \emph {et~al.}(2021)\citenamefont {Becker}, \citenamefont {Van~Manen}, \citenamefont {Haag}, \citenamefont {B{\"a}rlocher}, \citenamefont {Li}, \citenamefont {B{\"o}rsing}, \citenamefont {Curtis}, \citenamefont {Serra-Garcia},\ and\ \citenamefont {Robertsson}}]{becker2021broadband}%
  \BibitemOpen
  \bibfield  {author} {\bibinfo {author} {\bibfnamefont {T.~S.}\ \bibnamefont {Becker}}, \bibinfo {author} {\bibfnamefont {D.-J.}\ \bibnamefont {Van~Manen}}, \bibinfo {author} {\bibfnamefont {T.}~\bibnamefont {Haag}}, \bibinfo {author} {\bibfnamefont {C.}~\bibnamefont {B{\"a}rlocher}}, \bibinfo {author} {\bibfnamefont {X.}~\bibnamefont {Li}}, \bibinfo {author} {\bibfnamefont {N.}~\bibnamefont {B{\"o}rsing}}, \bibinfo {author} {\bibfnamefont {A.}~\bibnamefont {Curtis}}, \bibinfo {author} {\bibfnamefont {M.}~\bibnamefont {Serra-Garcia}},\ and\ \bibinfo {author} {\bibfnamefont {J.~O.}\ \bibnamefont {Robertsson}},\ }\bibfield  {title} {\bibinfo {title} {Broadband acoustic invisibility and illusions},\ }\href@noop {} {\bibfield  {journal} {\bibinfo  {journal} {Science advances}\ }\textbf {\bibinfo {volume} {7}},\ \bibinfo {pages} {9627} (\bibinfo {year} {2021})}\BibitemShut {NoStop}%
\bibitem [{\citenamefont {Ahmed}\ \emph {et~al.}(2021)\citenamefont {Ahmed}, \citenamefont {Farhat}, \citenamefont {Zhang},\ and\ \citenamefont {Wu}}]{ahmed2021deterministic}%
  \BibitemOpen
  \bibfield  {author} {\bibinfo {author} {\bibfnamefont {W.~W.}\ \bibnamefont {Ahmed}}, \bibinfo {author} {\bibfnamefont {M.}~\bibnamefont {Farhat}}, \bibinfo {author} {\bibfnamefont {X.}~\bibnamefont {Zhang}},\ and\ \bibinfo {author} {\bibfnamefont {Y.}~\bibnamefont {Wu}},\ }\bibfield  {title} {\bibinfo {title} {Deterministic and probabilistic deep learning models for inverse design of broadband acoustic cloak},\ }\href@noop {} {\bibfield  {journal} {\bibinfo  {journal} {Physical Review Research}\ }\textbf {\bibinfo {volume} {3}},\ \bibinfo {pages} {013142} (\bibinfo {year} {2021})}\BibitemShut {NoStop}%
\bibitem [{\citenamefont {Cominelli}\ \emph {et~al.}(2022)\citenamefont {Cominelli}, \citenamefont {Quadrelli}, \citenamefont {Sinigaglia},\ and\ \citenamefont {Braghin}}]{cominelli2022design}%
  \BibitemOpen
  \bibfield  {author} {\bibinfo {author} {\bibfnamefont {S.}~\bibnamefont {Cominelli}}, \bibinfo {author} {\bibfnamefont {D.~E.}\ \bibnamefont {Quadrelli}}, \bibinfo {author} {\bibfnamefont {C.}~\bibnamefont {Sinigaglia}},\ and\ \bibinfo {author} {\bibfnamefont {F.}~\bibnamefont {Braghin}},\ }\bibfield  {title} {\bibinfo {title} {Design of arbitrarily shaped acoustic cloaks through partial differential equation-constrained optimization satisfying sonic-metamaterial design requirements},\ }\href@noop {} {\bibfield  {journal} {\bibinfo  {journal} {Proceedings of the Royal Society A}\ }\textbf {\bibinfo {volume} {478}},\ \bibinfo {pages} {20210750} (\bibinfo {year} {2022})}\BibitemShut {NoStop}%
\bibitem [{\citenamefont {Popa}\ \emph {et~al.}(2011)\citenamefont {Popa}, \citenamefont {Zigoneanu},\ and\ \citenamefont {Cummer}}]{Popa2011}%
  \BibitemOpen
  \bibfield  {author} {\bibinfo {author} {\bibfnamefont {B.-I.}\ \bibnamefont {Popa}}, \bibinfo {author} {\bibfnamefont {L.}~\bibnamefont {Zigoneanu}},\ and\ \bibinfo {author} {\bibfnamefont {S.~A.}\ \bibnamefont {Cummer}},\ }\bibfield  {title} {\bibinfo {title} {Experimental acoustic ground cloak in air},\ }\href@noop {} {\bibfield  {journal} {\bibinfo  {journal} {Phys. Rev. Lett.}\ }\textbf {\bibinfo {volume} {106}},\ \bibinfo {pages} {253901} (\bibinfo {year} {2011})}\BibitemShut {NoStop}%
\bibitem [{\citenamefont {Zigoneanu}\ \emph {et~al.}(2014)\citenamefont {Zigoneanu}, \citenamefont {Popa},\ and\ \citenamefont {Cummer}}]{zigoneanu2014three}%
  \BibitemOpen
  \bibfield  {author} {\bibinfo {author} {\bibfnamefont {L.}~\bibnamefont {Zigoneanu}}, \bibinfo {author} {\bibfnamefont {B.-I.}\ \bibnamefont {Popa}},\ and\ \bibinfo {author} {\bibfnamefont {S.~A.}\ \bibnamefont {Cummer}},\ }\bibfield  {title} {\bibinfo {title} {Three-dimensional broadband omnidirectional acoustic ground cloak},\ }\href@noop {} {\bibfield  {journal} {\bibinfo  {journal} {Nature materials}\ }\textbf {\bibinfo {volume} {13}},\ \bibinfo {pages} {352} (\bibinfo {year} {2014})}\BibitemShut {NoStop}%
\bibitem [{\citenamefont {Bi}\ \emph {et~al.}(2018)\citenamefont {Bi}, \citenamefont {Jia}, \citenamefont {Sun}, \citenamefont {Yang}, \citenamefont {Zhao},\ and\ \citenamefont {Yang}}]{bi2018experimental}%
  \BibitemOpen
  \bibfield  {author} {\bibinfo {author} {\bibfnamefont {Y.}~\bibnamefont {Bi}}, \bibinfo {author} {\bibfnamefont {H.}~\bibnamefont {Jia}}, \bibinfo {author} {\bibfnamefont {Z.}~\bibnamefont {Sun}}, \bibinfo {author} {\bibfnamefont {Y.}~\bibnamefont {Yang}}, \bibinfo {author} {\bibfnamefont {H.}~\bibnamefont {Zhao}},\ and\ \bibinfo {author} {\bibfnamefont {J.}~\bibnamefont {Yang}},\ }\bibfield  {title} {\bibinfo {title} {Experimental demonstration of three-dimensional broadband underwater acoustic carpet cloak},\ }\href@noop {} {\bibfield  {journal} {\bibinfo  {journal} {Applied Physics Letters}\ }\textbf {\bibinfo {volume} {112}},\ \bibinfo {pages} {223502} (\bibinfo {year} {2018})}\BibitemShut {NoStop}%
\bibitem [{\citenamefont {Quadrelli}\ \emph {et~al.}(2021{\natexlab{a}})\citenamefont {Quadrelli}, \citenamefont {Cazzulani}, \citenamefont {{La Riviera}},\ and\ \citenamefont {Braghin}}]{quadrelli2021}%
  \BibitemOpen
  \bibfield  {author} {\bibinfo {author} {\bibfnamefont {D.~E.}\ \bibnamefont {Quadrelli}}, \bibinfo {author} {\bibfnamefont {G.}~\bibnamefont {Cazzulani}}, \bibinfo {author} {\bibfnamefont {S.}~\bibnamefont {{La Riviera}}},\ and\ \bibinfo {author} {\bibfnamefont {F.}~\bibnamefont {Braghin}},\ }\bibfield  {title} {\bibinfo {title} {Acoustic scattering reduction of elliptical targets via pentamode near-cloaking based on transformation acoustics in elliptic coordinates},\ }\href@noop {} {\bibfield  {journal} {\bibinfo  {journal} {Journal of Sound and Vibration}\ }\textbf {\bibinfo {volume} {512}},\ \bibinfo {pages} {116396} (\bibinfo {year} {2021}{\natexlab{a}})}\BibitemShut {NoStop}%
\bibitem [{\citenamefont {Urzhumov}\ \emph {et~al.}(2010)\citenamefont {Urzhumov}, \citenamefont {Ghezzo}, \citenamefont {Hunt},\ and\ \citenamefont {Smith}}]{Urzhumov2010}%
  \BibitemOpen
  \bibfield  {author} {\bibinfo {author} {\bibfnamefont {Y.}~\bibnamefont {Urzhumov}}, \bibinfo {author} {\bibfnamefont {F.}~\bibnamefont {Ghezzo}}, \bibinfo {author} {\bibfnamefont {J.}~\bibnamefont {Hunt}},\ and\ \bibinfo {author} {\bibfnamefont {D.~R.}\ \bibnamefont {Smith}},\ }\bibfield  {title} {\bibinfo {title} {Acoustic cloaking transformations from attainable material properties},\ }\href@noop {} {\bibfield  {journal} {\bibinfo  {journal} {New J. Phys.}\ }\textbf {\bibinfo {volume} {12}},\ \bibinfo {pages} {073014} (\bibinfo {year} {2010})}\BibitemShut {NoStop}%
\bibitem [{\citenamefont {Gao}\ and\ \citenamefont {Wu}(2016)}]{gao2016non}%
  \BibitemOpen
  \bibfield  {author} {\bibinfo {author} {\bibfnamefont {P.}~\bibnamefont {Gao}}\ and\ \bibinfo {author} {\bibfnamefont {L.}~\bibnamefont {Wu}},\ }\bibfield  {title} {\bibinfo {title} {Non-singular acoustic cloak derived by the ray tracing method with rotationally symmetric transformations},\ }\href@noop {} {\bibfield  {journal} {\bibinfo  {journal} {Proceedings of the Royal Society A: Mathematical, Physical and Engineering Sciences}\ }\textbf {\bibinfo {volume} {472}},\ \bibinfo {pages} {20150348} (\bibinfo {year} {2016})}\BibitemShut {NoStop}%
\bibitem [{\citenamefont {Cominelli}(2025)}]{cominelli2025non}%
  \BibitemOpen
  \bibfield  {author} {\bibinfo {author} {\bibfnamefont {S.}~\bibnamefont {Cominelli}},\ }\bibfield  {title} {\bibinfo {title} {Non-dispersive graded impedance acoustic lenses},\ }\href@noop {} {\bibfield  {journal} {\bibinfo  {journal} {arXiv preprint arXiv:2501.00821}\ } (\bibinfo {year} {2025})}\BibitemShut {NoStop}%
\bibitem [{\citenamefont {Dupont}\ \emph {et~al.}(2016)\citenamefont {Dupont}, \citenamefont {Guenneau}, \citenamefont {Kimmoun}, \citenamefont {Molin},\ and\ \citenamefont {Enoch}}]{dupont2016cloaking}%
  \BibitemOpen
  \bibfield  {author} {\bibinfo {author} {\bibfnamefont {G.}~\bibnamefont {Dupont}}, \bibinfo {author} {\bibfnamefont {S.}~\bibnamefont {Guenneau}}, \bibinfo {author} {\bibfnamefont {O.}~\bibnamefont {Kimmoun}}, \bibinfo {author} {\bibfnamefont {B.}~\bibnamefont {Molin}},\ and\ \bibinfo {author} {\bibfnamefont {S.}~\bibnamefont {Enoch}},\ }\bibfield  {title} {\bibinfo {title} {Cloaking a vertical cylinder via homogenization in the mild-slope equation},\ }\href@noop {} {\bibfield  {journal} {\bibinfo  {journal} {Journal of Fluid Mechanics}\ }\textbf {\bibinfo {volume} {796}},\ \bibinfo {pages} {R1} (\bibinfo {year} {2016})}\BibitemShut {NoStop}%
\bibitem [{\citenamefont {Chen}\ \emph {et~al.}(2017)\citenamefont {Chen}, \citenamefont {Zheng}, \citenamefont {Liu}, \citenamefont {Bi}, \citenamefont {Sun}, \citenamefont {Xiang}, \citenamefont {Yang},\ and\ \citenamefont {Hu}}]{chen2017broadband}%
  \BibitemOpen
  \bibfield  {author} {\bibinfo {author} {\bibfnamefont {Y.}~\bibnamefont {Chen}}, \bibinfo {author} {\bibfnamefont {M.}~\bibnamefont {Zheng}}, \bibinfo {author} {\bibfnamefont {X.}~\bibnamefont {Liu}}, \bibinfo {author} {\bibfnamefont {Y.}~\bibnamefont {Bi}}, \bibinfo {author} {\bibfnamefont {Z.}~\bibnamefont {Sun}}, \bibinfo {author} {\bibfnamefont {P.}~\bibnamefont {Xiang}}, \bibinfo {author} {\bibfnamefont {J.}~\bibnamefont {Yang}},\ and\ \bibinfo {author} {\bibfnamefont {G.}~\bibnamefont {Hu}},\ }\bibfield  {title} {\bibinfo {title} {Broadband solid cloak for underwater acoustics},\ }\href@noop {} {\bibfield  {journal} {\bibinfo  {journal} {Physical Review B}\ }\textbf {\bibinfo {volume} {95}},\ \bibinfo {pages} {180104} (\bibinfo {year} {2017})}\BibitemShut {NoStop}%
\bibitem [{\citenamefont {Quadrelli}\ \emph {et~al.}(2021{\natexlab{b}})\citenamefont {Quadrelli}, \citenamefont {Casieri}, \citenamefont {Cazzulani}, \citenamefont {La~Riviera},\ and\ \citenamefont {Braghin}}]{quadrelli2021experimental}%
  \BibitemOpen
  \bibfield  {author} {\bibinfo {author} {\bibfnamefont {D.~E.}\ \bibnamefont {Quadrelli}}, \bibinfo {author} {\bibfnamefont {M.~A.}\ \bibnamefont {Casieri}}, \bibinfo {author} {\bibfnamefont {G.}~\bibnamefont {Cazzulani}}, \bibinfo {author} {\bibfnamefont {S.}~\bibnamefont {La~Riviera}},\ and\ \bibinfo {author} {\bibfnamefont {F.}~\bibnamefont {Braghin}},\ }\bibfield  {title} {\bibinfo {title} {Experimental validation of a broadband pentamode elliptical-shaped cloak for underwater acoustics},\ }\href@noop {} {\bibfield  {journal} {\bibinfo  {journal} {Extreme Mechanics Letters}\ ,\ \bibinfo {pages} {101526}} (\bibinfo {year} {2021}{\natexlab{b}})}\BibitemShut {NoStop}%
\bibitem [{\citenamefont {Ginsberg}(2018)}]{ginsberg2018acoustics}%
  \BibitemOpen
  \bibfield  {author} {\bibinfo {author} {\bibfnamefont {J.~H.}\ \bibnamefont {Ginsberg}},\ }\href@noop {} {\emph {\bibinfo {title} {Acoustics: A Textbook for Engineers and Physicists}}},\ Vol.~\bibinfo {volume} {2}\ (\bibinfo  {publisher} {Springer},\ \bibinfo {year} {2018})\BibitemShut {NoStop}%
\bibitem [{\citenamefont {Ammari}\ \emph {et~al.}(2018)\citenamefont {Ammari}, \citenamefont {Fitzpatrick}, \citenamefont {Gontier}, \citenamefont {Lee},\ and\ \citenamefont {Zhang}}]{ammari2018minnaert}%
  \BibitemOpen
  \bibfield  {author} {\bibinfo {author} {\bibfnamefont {H.}~\bibnamefont {Ammari}}, \bibinfo {author} {\bibfnamefont {B.}~\bibnamefont {Fitzpatrick}}, \bibinfo {author} {\bibfnamefont {D.}~\bibnamefont {Gontier}}, \bibinfo {author} {\bibfnamefont {H.}~\bibnamefont {Lee}},\ and\ \bibinfo {author} {\bibfnamefont {H.}~\bibnamefont {Zhang}},\ }\bibfield  {title} {\bibinfo {title} {Minnaert resonances for acoustic waves in bubbly media},\ }in\ \href@noop {} {\emph {\bibinfo {booktitle} {Annales de l'Institut Henri Poincar{\'e} C, Analyse non lin{\'e}aire}}},\ Vol.~\bibinfo {volume} {35}\ (\bibinfo {organization} {Elsevier},\ \bibinfo {year} {2018})\ pp.\ \bibinfo {pages} {1975--1998}\BibitemShut {NoStop}%
\end{thebibliography}%


\end{document}